\documentclass[aps,prl,reprint,amsmath,amssymb,superscriptaddress,showpacs]{revtex4-1}
\usepackage{color}
\usepackage{enumitem}
\usepackage{graphicx}
\usepackage[normalem]{ulem}
\usepackage{comment}
\usepackage{xcolor,cancel}

\newif\ifHighlitedChanges
\def\ifHighlitedChanges{\iftrue}
\def\ifHighlitedChanges{\iffalse}
\ifHighlitedChanges
  \def\EDITS#1{{\color{red}#1}}
  \def\STRIKE#1{{\color{red}\sout{#1}}}
	\def\EDITSS#1{{\color{blue}#1}}
\else
  \def\EDITS#1{#1}
  \def\STRIKE#1{\relax}
	\def\EDITSS#1{\relax}
\fi
\begin{document}

\bibliographystyle{apsrev}

\title{Electrothermal Transistor Effect and Cyclic Electronic Currents \\
in Multithermal Charge Transfer Networks}
\author{Galen T. Craven}
%\email{gcraven@sas.upenn.edu}
\affiliation{Department of Chemistry, University of Pennsylvania, Philadelphia, PA  19104, USA} 
\author{Abraham Nitzan}
%\email{anitzan@sas.upenn.edu}
\affiliation{Department of Chemistry, University of Pennsylvania, Philadelphia, PA  19104, USA} 
\affiliation{School of Chemistry, Tel Aviv University, Tel Aviv 69978, Israel}

%%%%%%%%%%%%%%%%%%%%%%%%%%%%%%%%%%%%%%%%%%%%%%%%%%%%%%%%%%%%%%%%%%%%%%%%%%%%%%%%%%

\begin{abstract}
A theory is developed to describe the coupled transport of energy 
and charge in networks of electron donor-acceptor sites which are seated in a \STRIKE{thermally nonhomogeneous }\EDITS{thermally heterogeneous} environment, where
the transfer kinetics are dominated by Marcus-type hopping rates.
It is found that the coupling of heat and charge transfer in such systems 
gives rise to
exotic transport phenomena 
which are absent in 
thermally homogeneous systems and
cannot be described by standard thermoelectric relations.
Specifically, the directionality and extent of 
thermal transistor amplification and 
cyclical electronic currents 
in a given network can be controlled by tuning the underlying temperature gradient in the system.
The application of these findings toward optimal control of multithermal currents
is illustrated on a paradigmatic nanostructure.
\vspace{0.22cm}
\begin{description}
%\item[PACS numbers]
%82.20.Db, 05.40.Ca, 05.45.-a, 34.10.+x
\item[Cite as]
G. T. Craven and A. Nitzan, \textit{Phys. Rev. Lett.} \textbf{118}, 207201 (2017)
\item[DOI]
 \href{dx.doi.org/10.1103/PhysRevLett.118.207201}{10.1103/PhysRevLett.118.207201}
\end{description}
\end{abstract}
%\pacs{82.20.-w, 34.10.+x, 73.63.-b, 44., 66.30.Xj, 64.60.aq}

   \maketitle

The coupling between electron transfer (ET) and transport and the underlying thermal environment is a long studied subject \cite{Nitzan2007jpcm}. Its manifestations in recent studies of transport in molecular junctions mostly focus on the weak electron-phonon coupling regime. Similarly, thermoelectric phenomena in such junctions \cite{Dubi2011}, where molecules connect between electrodes of different temperatures, are usually treated 
(with a few exceptions, e.g., Refs.~\cite{Walczak2007,Galperin2008,Wang2011,Ren2012,Tagani2012,Tagani2013,Perroni2014,Perroni1015,Koch2014,Zimbovskaya2016})
with electron-vibration interaction disregarded. This stands in contrast to electron transfer reactions in condensed molecular systems that are usually dominated by hopping between thermally equilibrated polaron-like states as described by Marcus theory
\cite{Marcus1956,Marcus1964,Nitzan2006chemical,Peters2015} (\STRIKE{note that }analogous kinetics in junction transport is known, 
mostly in so called redox molecular junctions \cite{Zhang2008,Migliore2013}). 
We have recently considered the latter type of electronic transport in \STRIKE{thermally nonhomogeneous }\EDITS{thermally heterogeneous} systems, where an electron hops between two sites of different local temperatures \cite{craven16c}. 
This study was motivated by recent advances in measurement and control of temperature differences on length scales comparable to those involved in molecular electron transfer processes \cite{Chang2006,Sadat2010,Malen2010,Tan2011,Lee2013,Kim2014}.
The corresponding ET rate was obtained as a modified multidimensional Marcus expression that depends on the local temperatures of the two sites. Furthermore, electron hopping was shown to to be accompanied by heat transfer between the sites whose magnitude depends on the temperature difference and on the reorganization (polaron formation) energies at the two sites.

Thermal inhomogeneity can develop spontaneously in driven nonequilibrium systems \cite{Hern2007,Jarzynski2015nature,craven15c,Donadio2015,Nitzan2016network} or
can be externally controlled as in a thermoelectric device. 
%In the latter case 
Considering such models,
several recent
theoretical studies have discovered interesting thermal transistor effects, whereupon the flux
between two sites can be controlled by the temperature on a third site 
\cite{Li2006,Ben-Abdallah2014,Joulain2016}.

%%%%%%%%%%%%%%%%%%%%%
\begin{figure}[t]
\includegraphics[width = 8.6cm,clip]{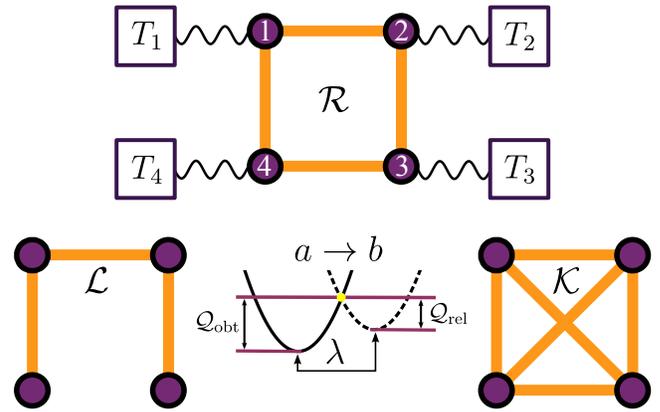}
\caption{\label{fig:Graph}
Network graphs for $\mathcal{R}$, $\mathcal{L}$, and $\mathcal{K}$ topologies.
Each node represents a donor-acceptor site.
As shown explicitly in the $\mathcal{R}$ graph, each site is in contact with an independent thermal bath.
The center panel shows a schematic of the energy surfaces $E_a$ (solid curve) and $E_b$ (dashed curve) 
and $\mathcal{Q}_\text{obt}$ and $\mathcal{Q}_\text{rel}$
for the state transition $a \to b$.
%\CR{The simplical complexes of each state in the corresponding network}
}
\end{figure}
%%%%%%%%%%%%%%%%%%%%%

In this Letter, 
we develop a theory to describe ET in complex networks
of donor-acceptor sites, where each site 
%contains 
is associated with
normal modes that are in
contact with an independent heat bath 
%at a temperature associated with the respective site.
at the local site temperature.
This is an idealization of the standard phenomenology of molecular electron transfer where the electronic processes is most strongly coupled to vibrations that are localized on the donor and acceptor sites.
%\EDITS{In the limit of strong electron-phonon coupling, electron transport is dominated by 
%hopping-type events and the Nobel-lauded theory 
%developed by Marcus has been shown to give quantitatively-accurate dynamical predictions
%while also describing the correct phenomenology \cite{Marcus1956,Marcus1964,Nitzan2006chemical,Peters2015}.}
In the limit of strong electron-phonon coupling, electron transport is dominated by 
hopping-type events and 
%the Nobel-lauded theory 
%developed by Marcus 
Marcus theory
gives the conceptual basis, and sometimes quantitative understanding, 
of ET reactions \cite{Marcus1956,Marcus1964,Nitzan2006chemical,Peters2015}.
The multithermal nature of the examined systems
arises because 
%each bath can take a different local temperature.
the donor and acceptor sites can be at different temperatures.
Purely vibrational heat transfer between sites is disregarded for simplicity, so in what follows we
focus on electron transfer and the associated heat transfer \cite{craven16c,matyushov16c}.
The development of complex network theory \cite{Schnakenberg1976} has significantly increased our understanding of the flow of charge, energy, and information in diverse types of systems \cite{Strogatz2001,Barabasi2002,Estrada2007,Stanley2011prl,Solomon2010,Nitzan2011ring,Nitzan2012ring,Savoie2014,Nitzan2014network,Solomon2016} and the present study makes it possible to consider electron and heat transport within such a framework and to study the consequence of their interdependence.

The donor-acceptor networks we consider consist of
$S$ sites,
where each site $s$, which has local temperature $T_s$, is associated with $N_s$ harmonic modes that are 
equilibrated with the 
thermal environment about that site. Specifically, we consider transitions between electronic states for which an excess electron is localized on different sites of this network.
In the Marcus picture \cite{Marcus1956} of ET, the localized occupation of electron density associated with electronic state $a$ of the network is described by the energy surface 
\begin{equation}
\label{eq:energysurf}
E_a(x_1,\ldots,x_N) =  E_a^{(0)}+\sum\limits^S_{{s}} \!\! \sum\limits_{j \in  \mathcal{M}^{(s)}} \frac{1}{2}k_j\left(x_j - \lambda^{(a)}_j\right)^2,
\end{equation}
where $E_a^{(0)}$ is an electronic energy origin of state $a$,
$x_j$ is the coordinate of the $j$th vibrational mode, and $\lambda^{(a)}_j$
is a shift in the equilibrium position of the $j$th mode.
%from some reference configuration 
%\EDITS{with energy $E_\text{ref} = \tfrac{1}{2} k_j x_j^2$
%which corresponds to the situation in which the current electronic state does not affect that mode.}
Both  $E_a^{(0)}$ and $\lambda^{(a)}_j$ are measured relative to some reference state for which $E_\text{ref}^{(0)}$ and $\lambda^{(\text{ref})}_j$ vanish (properties of this state do not affect the final results).
The total number of modes in the system is denoted by $N = \sum^S_s N_s$
and the group of modes associated with the $s$th site
is $\mathcal{M}^{(s)}$.
In a likely special case $\lambda^{(a)}_j = 0$
unless mode $j$ is
localized about the site on which the electron is placed in state $a$ (that is, unless $j \in  \mathcal{M}^{(s)}$ and
$a$ is the electronic state that corresponds to the electron occupying site $s$).
However, the form (\ref{eq:energysurf})
can represent more general situations where the modes localized about site $s$ respond to the electronic
occupation on a different site.
A transition between states $a$ and $b$ is associated with a
reorganization energy (assumed temperature independent) in the $j$th mode, $E^{(a,b)}_{\text{R}j}$,
and a total reorganization energy, $E^{(a,b)}_{\text{R}}$, which are given by: 
%$E^{(a,b)}_{\text{R}j}  = \frac{1}{2}k_j \Delta \lambda_j^2$ and $E^{(a,b)}_{\text{R}}  = \sum_j^N E^{(a,b)}_{\text{R}j}$.
\begin{equation}
E^{(a,b)}_{\text{R}j}  = \frac{1}{2}k_j \Big(\lambda_j^{(a)}-\lambda_j^{(b)}\Big)^2  \quad \text{and} \quad E^{(a,b)}_{\text{R}}  = \sum_j^N E^{(a,b)}_{\text{R}j}.
\end{equation}

The shifts $\lambda$, and hence the reorganization energies, of a system depend on the intersite distances and therefore the topology of the underlying connectivity network.
Shown in Fig.~\ref{fig:Graph}
are representative topologies for three typical connectivities: 
ring $\mathcal{R}$, linear $\mathcal{L}$,
and complete $\mathcal{K}$.
Associated with each network is an adjacency matrix
$\mathbf{A}$
where
$\text{A}_{ab} = 1$ if sites $a$ and $b$ are connected 
(and thus the electron can tunnel between sites)
and is zero otherwise \cite{Schnakenberg1976}.
The topology of $\mathbf{A}$ determines
which modes are responsive to electron localization on a particular site.
%Each site $s$ in the network, 
%and the modes associated with that site,
%are connected to a heat bath of temperature $T_s$ (see Fig.~\ref{fig:Graph}).
%\CR{This state transition is a simplical complex that depends on the network topology}
We next show that 
%multithermal temperature gradients
temperature differences 
between sites give rise to 
emergent and sometimes exotic thermal and electronic transport properties.

The electron transfer rate between any two sites in the network and the
heat transfer rate between the thermal baths involved in this transition can
be derived by adopting the formalism put forth in Ref.~\citenum{craven16c}
for bithermal electron hopping between two sites.
``Involved bath'' implies that the
harmonic modes that are thermalized by
%by a given 
this bath are sensitive to the electronic population of at least
one of the sites.
%\CR{, i.e., that the site is a member of the simplical complex corresponding to either state $a$ or state $b$}.
%The results given below are generalizations\cite{note1} to many sites and many
%modes per site of those derived in Ref.~\citenum{craven16c}.
Under standard transition state theory assumptions the
rate of the $a \to b \equiv a,b$ transition  can be expressed as
\begin{equation}
\label{eq:genrate}
k_{a,b} = \frac{1}{2} \Big\langle \mathcal{T}_{a,b} \, \dot{x}_\perp\Big\rangle P_{a,b}, 
\end{equation}
where $\mathcal{T}_{a,b}$ is the tunneling probability between states, 
$P_{a ,b}$ is the probability density about a transition surface (TS) separating the states,
both evaluated on the $E_a$ surface,
and $\dot{x}_\perp$
is the velocity in the direction normal to the TS \cite{Schiffer1995,Jonsson2001}.
The normal velocity $\dot{x}_\perp = \dot{\mathbf{x}}  \cdot \mathbf{\hat{u}_\perp}(\mathbf{x})$ where $\mathbf{\hat{u}_\perp}(\mathbf{x})$ is a unit vector normal to the TS at position $\mathbf{x}$ on this hypersurface.
The factor $\langle \mathcal{T}_{a,b} \, \dot{x}_\perp\rangle$ is a (multi)thermal average 
that depends on the temperature of each bath involved in the transition.
%The rate is a product of this factor and the probability $P_{a,b}$ to observe the system
%at the TS separating reactants and products.
The TS is determined by the requirement that a transition can only take place at nuclear configurations where electronic energy is
conserved.
For the $N$-dimensional paraboloid energy surface defined by (\ref{eq:energysurf}),
the TS separating states is an $(N-1)$-dimensional transition state hypersurface  
which is the locus of mode configurations 
where $E_a = E_b$, defined by
$g_\text{c}(x_1,\ldots,x_N) = E_b(x_1,\ldots,x_N)-E_a(x_1,\ldots,x_N)$.
%The condition for an electron to transfer from state $a$ to state $b$ is $E_a = E_b$, i.e., the transfer can occur only
%when each state is at the same energy. This is a consequence of energy conservation.

In the adiabatic limit of Marcus rate theory $\mathcal{T}_{a,b}=1$
and the pre-exponential factor 
is proportional to $\left\langle \dot{x}_\perp\right\rangle$ \cite{note1}.
%The velocity vector of the system
%$\dot{\mathbf{x}} = \left\{\dot{x}_1,\ldots,\dot{x}_N\right\}$
%has a scalar component in the direction of the normal
%$\dot{x}_\perp(\dot{x}_1,\ldots,\dot{x}_N) = \dot{\mathbf{x}}  \cdot \mathbf{\hat{u}_\perp} = \dot{x}_1 n_1 + \ldots + \dot{x}_N n_N$,
%where $\mathbf{\hat{u}_\perp} = \left\{n_1,\ldots,n_N\right\}$ is a unit vector normal to the TSHP.
%The expectation value of $\dot{x}_\perp$
%can be evaluated using the Heaviside function to 
%constrain the expectation integral over the space in which $\dot{x}_\perp \geq 0$ \cite{note1}:
%\begin{equation}
%\label{eq:vel}
%\left\langle \dot{x}_\perp \right\rangle = \frac{\displaystyle\int_{\mathbb{R}^N} \dot{x}_\perp\;\Theta\big(\dot{x}_\perp\big)\prod_{s}^S \!\!\prod_{{j \in  \mathcal{M}^{(s)}}} \!\! \exp\left[{-\beta_s\frac{1}{2}m_j \dot{x}_j^2}\right]   d\dot{x}_j}
%{\displaystyle\int_{\mathbb{R}^N} \Theta\big(\dot{x}_\perp\big)\prod_{s}^S \!\!\prod_{{j \in  \mathcal{M}^{(s)}}} \!\!  \exp\left[{-\beta_s \frac{1}{2}m_j \dot{x}_j^2}\right] d\dot{x}_j},
%\end{equation}
%where $\beta_s = 1/k_\text{B} T_s$ and $m_j$ is the mass associated with the $j$th mode.
%Evaluating the integrals in Eq.~(\ref{eq:vel}) yields
%\begin{equation}
%\label{eq:velformTsitesnormal}
%\left\langle \dot{x}_\perp \right\rangle = \sqrt{ \displaystyle \frac{2 k_\text{B}}{\pi}  \sum_{s}^S T_s \!\!\! \sum_{{j \in  \mathcal{M}^{(s)}}} \!\!\! n_j^2 \Big/ \displaystyle  m_j}.
%\end{equation}
In the nonadiabatic limit \cite{notenonadiabatic},
%\EDITS{which is more appropriate to the present consideration,}
 $\mathcal{T}_{a,b}$ can be approximated 
%using  
by the corresponding limit of
the Landau-Zener expression \cite{Thompson1998multi,Nitzan2006chemical,Varganov2016}
which is evaluated in the direction normal to the TS \cite{Marks1992}.
In this limit, $\mathcal{T}_{a,b} \propto 1/ \dot{x}_\perp$, and the expectation value
$\left\langle \mathcal{T}_{a,b}\, \dot{x}_\perp\right\rangle$ does not depend on the normal velocity \cite{note4}.
For the $a \to b$ transition, the multithermal probability to be on the TS is:
\begin{equation}
\begin{aligned}
\label{eq:kabflux} 
P_{a,b} &=  \int_{\mathbb{R}^N} \left| \nabla g_\text{c} \right| \delta\big(g_\text{c}(x_1,\ldots,x_N)\big)  \\[0ex]
& \quad \times  \prod_{s}^S \!\!\prod_{{j \in  \mathcal{M}^{(s)}}} \!\! \exp\left[{{-\beta_s \frac{k_j}{2} \left(x_j-\lambda^{(a)}_j\right)^2}}\right]  dx_j\\[0ex]
& \Bigg/  \!\!\int_{\mathbb{R}^N} \prod_{s}^S \!\!\prod_{{j \in  \mathcal{M}^{(s)}}} \!\! \exp\left[{{-\beta_s \frac{k_j}{2}\left(x_j-\lambda^{(a)}_j\right)^2}}\right] dx_j,
\end{aligned}
\end{equation}
where $\beta_s = 1/k_\text{B} T_s$ with $k_\text{B}$ being the Boltzmann constant.
The $\delta$-function in (\ref{eq:kabflux}) constrains the integration over the vibrational subspace in which $E_a = E_b$ and
the gradient magnitude $\left| \nabla g_\text{c} \right| =  \sqrt{\sum_j^N 2 k_j E^{(a,b)}_{\text{R}j}}$
gives a precise definition to this constraint \cite{Hartmann2011deltafunction}.
Evaluating %and the coarea formula \cite{note1} to evaluate 
the integrals in Eq.~(\ref{eq:kabflux}) we obtain   
\begin{equation}
\begin{aligned}
\label{eq:probdens}
&P_{a,b} =   \displaystyle  \sqrt{ \sum_{j}^N k_j E^{(a,b)}_{\text{R}j}  \Bigg/   2 \pi k_\text{B} \sum_{s}^S T_s \!\!\! \sum_{{j \in  \mathcal{M}^{(s)}}} \!\!\! E^{(a,b)}_{\text{R}j}  } \\[0ex]
&\times \exp\left[-  \displaystyle  \left(E_{ba}+E^{(a,b)}_\text{R}\right)^2 \Bigg/ \displaystyle 4  k_\text{B} \sum_{s}^S T_s \!\!\! \sum_{{j \in  \mathcal{M}^{(s)}}} \!\!\! E^{(a,b)}_{\text{R}j}\right],
\end{aligned}
\end{equation}
which expresses the probability density on the reaction path in terms of the temperature of each bath, and the reorganization energy and reaction free energy
$E_{ba} = -E_{ab} = E^{(0)}_b - E^{(0)}_a$ of the respective transition.
In the uniform temperature limit, combining Eqs.~(\ref{eq:genrate}) and (\ref{eq:probdens}) recovers the Marcus rate \cite{Marcus1956,Marcus1964,Nitzan2006chemical,Peters2015}.
%which has been shown to successfully describe many aspects of ET in molecular systems .}

With the multidimensional-multithermal transition rate $k_{a,b}$
now derived, the kinetic equations for the occupation probability $\mathcal{P}$
of each state can be constructed.
For a reaction network (see Fig.~\ref{fig:Graph}) with adjacency matrix $\mathbf{A}$ these equations take the form
\begin{equation}
\frac{d \mathcal{P}_a}{dt} = \sum_b \text{A}_{ba} k_{b,a}  \mathcal{P}_b(t)- \text{A}_{ab} k_{a,b} \mathcal{P}_a(t),
\end{equation}
for each state $a$.
At steady-state $d \mathcal{P}_a / dt = 0 \: \forall \, a$, which implies that 
the net electronic flux between sites vanishes.
%Of significance is that in these multithermal systems the
%occupancy probability currents 
%$\dot{\mathcal{P}}$,
%which are proportional to the electric current,
%are coupled with the transport of heat.
%The electronic contribution to heat transport 
%has previously been characterized for a bithermal two-site electron hopping
%system \cite{craven16c},
%and this formalism can be extended to electron hopping in networks of donor and acceptor sites 
%across complex thermal gradients using the transfer rates and kinetic equations derived here.

%In multithermal-multidimensional ET networks, each reaction path leading to a state transition proceeds through a point
Next consider the heat transfer associated with a given electron transfer step \cite{craven16c,craven17a}.
In Marcus theory, the nuclear motion leading to the $a \to b$ transition proceeds through a point
$\boldsymbol{x}^\text{TS}$
on the TS, and the corresponding heat transferred into 
a specific bath during the transition is 
$\mathcal{Q}^{(a,b)}(\boldsymbol{x}^\text{TS}) = -\mathcal{Q}^{(a)} _\text{obt}(\boldsymbol{x}^\text{TS}) + \mathcal{Q}^{(b)}_\text{rel}(\boldsymbol{x}^\text{TS}),$
where the first term is the heat 
%released by 
obtained from
the bath during the ascent to $\boldsymbol{x}^\text{TS}$
on the $E_a$ surface and the second term is the heat
%absorbed by
released to
the bath during the 
descent to equilibrium on the $E_b$ surface (see Fig.~\ref{fig:Graph}).
Using the energy surfaces of Eq.~(\ref{eq:energysurf}),
the contribution of mode $j$ to the net heat exchange with the bath associated with it during the $a \to b$ transition is given by
%the heat transfer term for the bath associated with the $j$th mode can be written as
\begin{equation}
   \mathcal{Q}^{(a,b)}_j = -\frac{1}{2} k_j \left[x^\text{TS}_j -\lambda_j^{(a)}\right]^2 + \frac{1}{2} k_j \left[x^\text{TS}_j -\lambda_j^{(b)}\right]^2.
\end{equation}
%The expectation value for the heat transferred during the $a \to b$ transition is
This quantity should be averaged over the probability to reach the particular configuration on the TS when coming from  the $a$ side:
\begin{equation}
\label{eq:heatexpec}
  \Big\langle \mathcal{Q}^{(a,b)}_j \Big\rangle = \int_{\mathbb{R}^N}    \mathcal{Q}^{(a,b)}_j\, P^\ddag_{a,b}(x_1,\ldots,x_N)  \,dx_1 \ldots dx_N,
\end{equation}
where $P^\ddag_{a,b}$ is the probability density on the TS when the system is in state $a$ \cite{note1}:
\begin{equation}
\begin{aligned}
\label{eq:probdensTSHP}
P^\ddag_{a,b} &= \displaystyle \delta\left(g_\text{c}\right) \prod_{s}^S \!\!\prod_{{j \in  \mathcal{M}^{(s)}}} \!\! \exp\left[{{-\beta_s \frac{1}{2}k_j \left(x_j - \lambda^{(a)}_j\right)^2}}\right] \\
\Bigg/ &\int_{\mathbb{R}^N} \!\delta\left(g_\text{c}\right)  \prod_{s}^S \!\! \prod_{{j \in  \mathcal{M}^{(s)}}} \!\! \exp\left[{{-\beta_s \frac{1}{2}k_j \left(x_j - \lambda^{(a)}_j\right)^2}}\right] dx_j.
\end{aligned}
\end{equation}
Evaluating the expectation integral gives
\begin{equation}
\label{eq:heat}
\Big\langle \mathcal{Q}^{(a,b)}_j\Big\rangle = 
\frac{ \displaystyle E^{(a,b)}_{\text{R}j} \!\! \left[E_{ab} T_j + \! \sum_{k \neq j}^N  E^{(a,b)}_{\text{R}k} \! \big(T_k-T_j\big) \right]}{\displaystyle  \sum_{k}^N T_{k} E^{(a,b)}_{\text{R}k}}
\end{equation}
where $T_j = T_s$ if $ j \in \mathcal{M}^{(s)}$.
%For a specific site $s$ 
The total heat transferred to the thermal environment of site $s$
during the $a \to b$ transition is 
the sum of contributions over all modes associated with that site:
%\begin{equation}
%\label{eq:heatsum}
%\Big\langle \mathcal{Q}^{(a,b)}_s\Big\rangle =  \sum_{{j \in  \mathcal{M}^{(s)}}} \!\! \Big\langle \mathcal{Q}^{(a,b)}_j \Big\rangle .
%\end{equation}
%Combining Eqs.~(\ref{eq:heat}) and (\ref{eq:heatsum}) yields
\begin{equation}
\begin{aligned}
\label{eq:heatsite}
&\Big\langle \mathcal{Q}^{(a,b)}_s\Big\rangle = \\
&\frac{ \displaystyle \sum_{{j \in  \mathcal{M}^{(s)}}} \!\!\! E^{(a,b)}_{\text{R}j} \!\! \left[E_{ab} \, T_s + \! \sum_{q \neq s}^S  \sum_{{k \in  \mathcal{M}^{(q)}}} \!\!\! E^{(a,b)}_{\text{R}k}  \big(T_q-T_s\big) \right]}{\displaystyle  \sum_{q}^S T_{q} \!\!\! \sum_{{k \in  \mathcal{M}^{(q)}}} \!\!\! E^{(a,b)}_{\text{R}k}}.
\end{aligned}
\end{equation}
Here, the term proportional to $E_{ab}$ expresses the heat released to/taken from the baths from the free energy difference between these electronic states, while the term proportional to $(T_q-T_s)$ is the actual heat transfer between baths $q$ and $s$ associated with the $a \to b$ transition.
%which illustrates that the amount of heat transferred depends on the %temperature differences between the site $s$ and all other sites with modes %that are affected by the respective transition.

%%%%%%%%%%%%%%%%%%%%%
\begin{figure}[t]
\includegraphics[width = 8.6cm,clip]{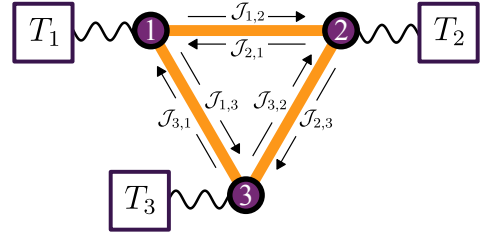}
\caption{\label{fig:Graph3}
Network graph for a three-site ring ($\mathcal{R}_3$) network.
}
\end{figure}
%%%%%%%%%%%%%%%%%%%%%

The heat current into  the thermal environment of site $s$ is 
\begin{equation}
\frac{d \mathcal{Q}_s}{d t} = \sum_{a,b} \text{A}_{ab} k_{a,b} \mathcal{P}_a(t) \Big\langle \mathcal{Q}^{(a,b)}_s\Big\rangle.
\end{equation}
Note that while $\mathcal{P}_a(t)\geq 0\,\forall\,t$, the expectation value of $\mathcal{Q}^{(a,b)}_s$ and the heat current $\dot{\mathcal{Q}}_s$ can be either positive or negative implying that the corresponding bath absorbs or releases energy, respectively.
In unithermal systems
at steady-state the heat currents vanish ($\dot{\mathcal{Q}}_s = 0\,\forall\,s$),
while in contrast, 
for multithermal systems, $\langle \mathcal{Q}_s\rangle \neq 0$ and
thus $\dot{\mathcal{Q}}_s \neq 0$.
%These currents are multithermal because Eq.~(\ref{eq:heat}) depends on the temperature difference with
%every site that feels the respective transition.
Note that
even as the occupation probabilities approach electronic quasi-equilibrium where the net electronic currents are zero,
the net flow of heat does not vanish.
This phenomenon is not a standard thermoelectric effect
and reveals a novel pathway for energy transport in multithermal charge transfer networks.

%%%%%%%%%%%%%%%%%%%%%
\begin{figure}
\includegraphics[width = 8.6cm,clip]{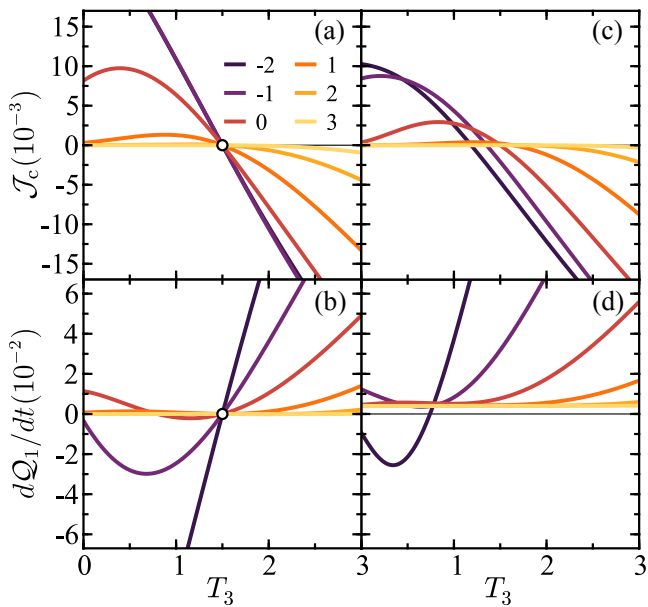}
\caption{\label{fig:FluxandCurrent} Cyclic flux $\mathcal{J}_\text{c}$ (top) and heat current $d \mathcal{Q}_1 / d t$ (bottom) in the nonadiabatic limit as functions of $T_3$
for a three-site $\mathcal{R}_3$ network.
Each curve corresponds to the respective value of $E_3^{(0)}$ marked in the legend.
In each panel: $E_1^{(0)} = -4$, $E_2^{(0)} = 0$, and 
$T_2 = 3/2$ with (a)-(b) $T_1$ = 3/2 and (c)-(d) $T_1=3/4$. 
The circular markers denote unithermal points.
%The circular markers denote points where $T_1 = T_2 = T_3$. 
All reorganization energies are $E_{\text{R}j} = 1/2$ for each mode involved in a particular
a transition and zero otherwise.
All quantities are shown in reduced units \cite{note4,note2}.
}
\end{figure}
%%%%%%%%%%%%%%%%%%%%%

%In multithermal systems with cycles (closed loops) in the connectivity network 
%we also observe a novel transport phenomena that
%is absent in unithermal systems, namely, 
%the unidirectional electronic flux $\mathcal{J}_{a,b}= k_{a,b} \mathcal{P}_a$ 
%through a specific connection $a \to b$ is not equal to the
%flux through the reverse connection $b \to a$, even at steady-state.
%In the unithermal case, all unidirectional fluxes through a specific connection are equal:
%$\mathcal{J}_{a,b} = \mathcal{J}_{b,a}$ (see Fig.~\ref{fig:Graph3}).
%However, this is not the case for networks of sites that are part of cycles in multithermal systems where
%an electric current can be induced around a ring structure by alteration of the bath temperatures.

%To illustrate the use of thermally-induced cyclical electronic currents in ET networks,
%a simple three-site ring ($\mathcal{R}_3$) network can be applied (see Fig.~\ref{fig:Graph3}).
Another interesting behavior that is observed in multithermal networks with closed loops 
is the persistence of steady-state net electronic bond currents, 
$\mathcal{J}_{a,b} - \mathcal{J}_{b,a}= k_{a,b} \mathcal{P}_a-k_{b,a} \mathcal{P}_b$, 
i.e., the breaking of detailed balance,
and the formation of cyclical current loops that obviously vanish in full equilibrium 
where detailed balance is maintained.
An example is seen in the three-site ring ($\mathcal{R}_3$) shown in  Fig.~\ref{fig:Graph3}. 
%We emphasize...
We emphasize this simple system because of
its experimental realizability and its general applications in molecular electronics and devices \cite{Li2012,Joulain2016,Allahverdyan2016}, but note that other
more complex networks can also be analyzed using the developed theory \cite{note5}.
In a multithermal $\mathcal{R}_3$ network, 
the direction and magnitude of the cyclic flux $\mathcal{J}_\text{c} = \mathcal{J}_{1,2}- \mathcal{J}_{2,1}= \mathcal{J}_{2,3}  - \mathcal{J}_{3,2} =  \mathcal{J}_{3,1}  - \mathcal{J}_{1,3}$ can be altered by tuning the temperature $T_3$.
This is illustrated in Fig.~\ref{fig:FluxandCurrent}(a) where 
%the donor and acceptor 
the donor and acceptor sites 1,2 are at the same temperature while variation of the temperature on another site, $T_3$, 
determines the direction and magnitude of the cyclical current.
We find that this thermally-induced current persists, 
except in the case that the electronic occupation energy $E_a^{(0)}$ is the same for all sites involved in the cycle.
Note that at the unithermal point in Figs.~\ref{fig:FluxandCurrent}(a)-(b), 
$\mathcal{J}_{a,b} = \mathcal{J}_{b,a}$ for every connection, 
so $\mathcal{J}_\text{c} =0$.
This is simply a statement that with no temperature gradient
there is no heat current or cyclic electron flux.
Similar trends are also observed in Figs.~\ref{fig:FluxandCurrent}(c)-(d) 
for networks where the local temperature of each site is different.
By comparing the temperatures at which $\mathcal{J}_\text{c} =0$ in Fig.~\ref{fig:FluxandCurrent}(c)
with the heat currents at the same temperature in Fig.~\ref{fig:FluxandCurrent}(d)
it can be seen that even when the electron flux vanishes there is still a net heat flow between baths.
At full equilibrium, which is achieved in the uniform temperature limit, all electronic and heat fluxes vanish.

%%%%%%%%%%%%%%%%%%%%%
\begin{figure}
\includegraphics[width = 8.6cm,clip]{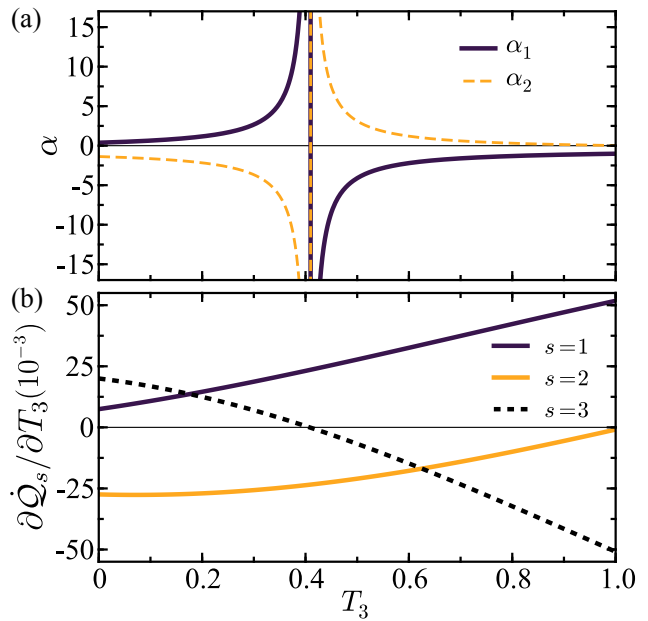}
\caption{\label{fig:Amp} (a) Amplification factor $\alpha$ and (b) heat current derivatives $\partial \dot{\mathcal{Q}}_s/\partial T_3$ as functions of $T_3$ for a 
multithermal $\mathcal{R}_3$ system with $T_1 = 3/4$, $T_2 = 3/2$, and $E_{\text{R}j} = 1/2  \: \forall \, j$ over each state transition.
In both panels $E_1^{(0)} = -4$ and $E_2^{(0)} = E_3^{(0)} = 0$.
}
\end{figure}
%%%%%%%%%%%%%%%%%%%%%

The theory of coupled  electron and heat transfer developed in this Letter can also be applied to elucidate the electronic contribution to transport phenomena in
thermal transistors \cite{Li2006,Ben-Abdallah2014,Joulain2016},
which are recently 
studied model devices
%developed devices 
that can be used to control and amplify heat flow.
%The magnitude of amplification in thermal transistors 
%can be quantified by the factor 
Following Ref.~\citenum{Li2006},
we quantify the magnitude of amplification in thermal transistors by the factor
$\alpha_s = \partial \dot{\mathcal{Q}}_s/\partial \dot{\mathcal{Q}}_3:s \in \left\{1,2\right\}$,
which measures the effect of pumping heat into site 3 
on the heat current between sites 1 and 2.
If $|\alpha_s|>1$ a transistor effect is present.
In the three-site $\mathcal{R}_3$ system,
variation of the heat current $\dot{\mathcal{Q}}_3$
by alteration of $T_3$
can give rise to significant amplifications,
as shown in Fig.~\ref{fig:Amp}(a). %\cite{note9}.
The reason for the electrothermal transistor effect is shown in Fig.~\ref{fig:Amp}(b),
where $\partial \dot{\mathcal{Q}}_3/\partial T_3 \to 0$  at $T_3 \approx 0.4$ 
while $|\partial \dot{\mathcal{Q}}_s/\partial T_3| \gg \partial \dot{\mathcal{Q}}_3/\partial T_3$ for $s \in \left\{1,2\right\}$ 
as this limit is approached.
This aligns with the region of amplification shown in Fig.~\ref{fig:Amp}(a).

We have shown that in nanoscale systems where localized modes are in contact with environments at different temperatures, 
complex nonlinearities in the thermal gradient can induce currents which are characterized by multiple temperatures. 
A theory has been developed to unify the description of heat and charge transfer in these systems
with multithermal temperature gradients between donor-acceptor states.
This work provides a bridge connecting theories of electron transfer, heat transport, and
thermoelectricity
in systems where electron transport is dominated by intersite hopping, and will be useful in the design of electronic and thermoelectric devices that operate in this limit.

In regimes where the magnitude of heat conduction due to electron transport 
dominates over the magnitude of conduction from phonons,
the developed theory will be directly applicable.
Otherwise, a complete picture of the conduction process
will require a theory that considers both electrothermal and phononic heat transport, and their coupling.
Examination of thermopower, efficiency, and their relation to electrothermal
transport in molecular junctions
(and other complex donor/acceptor networks with molecule-metal and molecule-semiconductor interfaces) in the 
phonon-assisted hopping limit of electronic transport
%limit of strong electron-phonon coupling 
is a potential application of 
%this 
the present
theory \cite{Jia2016,Gagorik2017,craven17a}.
Generalizations that go beyond the semiclassical Marcus treatment are obviously needed and will be taken on in future work.
%and could give rise to the design of new technologies with continuing thermal transport 
%even for vanishing thermoelectric coefficients.

AN is supported by the Israel Science Foundation, the US-Israel Binational Science Foundation and the University of Pennsylvania.
%This research has been supported by 
%discretionary funds from the Department of Chemistry at the University of Pennsylvania.
%%%%%%%%%%%%%%%%%%%%%%

%\bibliographystyle{apsrev}
\bibliography{j,electron-transfer,networks,craven,osc-bar,pensphere,nonequilibrium,c3_Arxiv}

\begin{thebibliography}{61}
\bibitem[{not({\natexlab{b}})}]{notenonadiabatic}
\bibinfo{note}{For experimental setups in which sites of appreciably different
  temperatures are relatively far from each other, implying relatively weak
  electronic coupling, the nonadiabatic limit will be the most pertinent.}
\bibitem[{not({\natexlab{c}})}]{note4}
\bibinfo{note}{In the nonadiabatic limit a coupling constant between diabatic
  surfaces: $H_{a,b}=0.01\,\text{eV}$ has been used in all calculations}.
\bibitem[{not({\natexlab{d}})}]{note2}
\bibinfo{note}{Characteristic units are: $\epsilon = 0.02\,\text{eV}$ for
  energy, $\tau = 0.5\,\text{ps}$ for time, and $\sigma = 1\,\text{nm}$ for
  length. Using these units, a reduced temperature of $k_\text{B} T/\epsilon
  \approx1.3$ corresponds to $T = 300\,\text{K}$.}
\end{thebibliography}
\end{document}